\documentclass[twocolumn,showpacs,amssymb,aps,superscriptaddress,pre]{revtex4}
\usepackage{graphicx}
\usepackage{dcolumn}
\usepackage{bm}

\begin{document}

\title{Magnetic properties of the Hubbard model on kagome stripes}

\author{O. Derzhko}
\affiliation{Institute for Condensed Matter Physics, NAS of Ukraine, 1 Svientsitskii Street, L'viv-11, 79011, Ukraine}
\affiliation{Max-Planck-Institut f\"{u}r  Physik Komplexer Systeme, N\"{o}thnitzer Stra\ss e 38, 01187 Dresden, Germany}

\author{M. Maksymenko}
\affiliation{Institute for Condensed Matter Physics, NAS of Ukraine, 1 Svientsitskii Street, L'viv-11, 79011, Ukraine}

\author{J. Richter}
\affiliation{Institut f\"{u}r Theoretische Physik, Universit\"{a}t Magdeburg, P.O. Box 4120, 39016 Magdeburg, Germany}

\author{A. Honecker}
\affiliation{Institut f\"{u}r Theoretische Physik, Georg-August-Universit\"{a}t G\"{o}ttingen,
             Friedrich-Hund-Platz 1, 37077 G\"{o}ttingen, Germany}

\author{R. Moessner}
\affiliation{Max-Planck-Institut f\"{u}r  Physik Komplexer Systeme, N\"{o}thnitzer Stra\ss e 38, 01187 Dresden, Germany}

\date{\today}

\pacs{71.10.-w}

\keywords{Hubbard model, kagome stripes, ferromagnetism}

\begin{abstract}
We consider the one-orbital $N$-site repulsive Hubbard model on two kagome-like chains,
both of which yield a completely dispersionless (flat) one-electron band.
Using exact many-electron ground states in the subspaces with $n\le n_{\max}$
($n_{\max}\propto N$)
electrons,
we calculate the square of the total spin in the ground state
to discuss magnetic properties of the models.
We have found that although for $n<n_{\max}$ the ground states contain fully polarized states,
there is no finite region of electron densities $n/{\cal{N}} <1$
(${\cal{N}}=N/3$ or ${\cal{N}}=N/5$)
where ground-state ferromagnetism survives
for ${\cal{N}}\to\infty$.
\end{abstract}

\maketitle


Mechanisms of ferromagnetism in the itinerant electron systems continue to be of great interest \cite{tasaki}.
In the early 1990s Mielke and Tasaki proposed several Hubbard models
with a unique saturated ferromagnetic ground state \cite{mielke1,tasaki1}.
This model has an important feature that the lowest-energy one-electron band is completely flat (dispersionless)
and, as a result, the corresponding one-electron states can be localized on a
small part of the lattice (trapping cell).
On the other hand,
exact ground states of frustrated quantum spin antiferromagnets have been discovered \cite{locmag}.
These states consist of independent (i.e., isolated) localized magnons located within trapping cells
and they dominate the low-temperature properties of the antiferromagnetic Heisenberg model in a strong magnetic field.
Interestingly,
although the flat-band repulsive Hubbard model far below half filling
and the highly frustrated Heisenberg antiferromagnet in a strong magnetic field
present different physics,
the mathematical structure for their description may be quite similar \cite{locmag,dr06,hr,dhr07,drhmm}.

\begin{figure}[htp]
\begin{center}
\includegraphics[angle = 0, width = 0.35\textwidth]{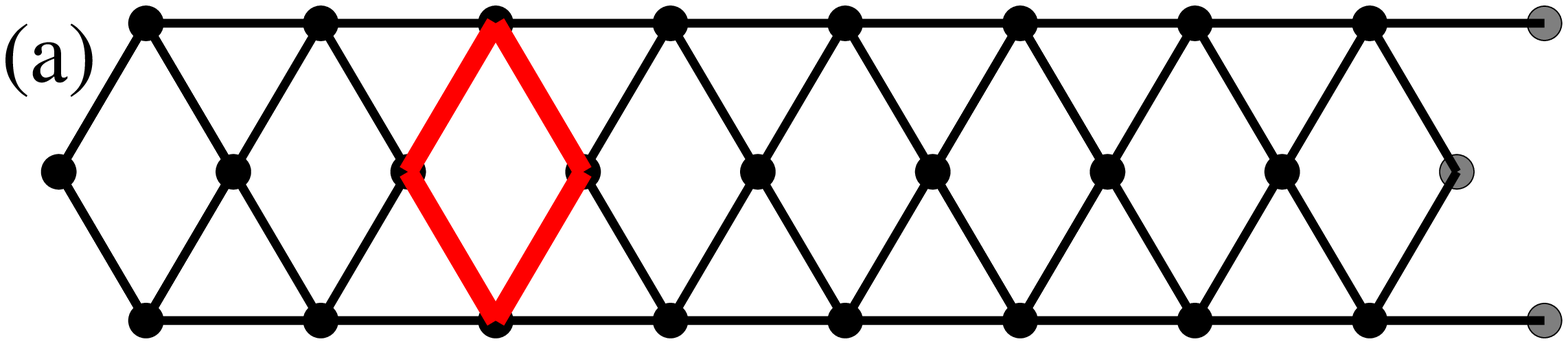}
\\
\vspace{2mm}
\includegraphics[angle = 0, width = 0.35\textwidth]{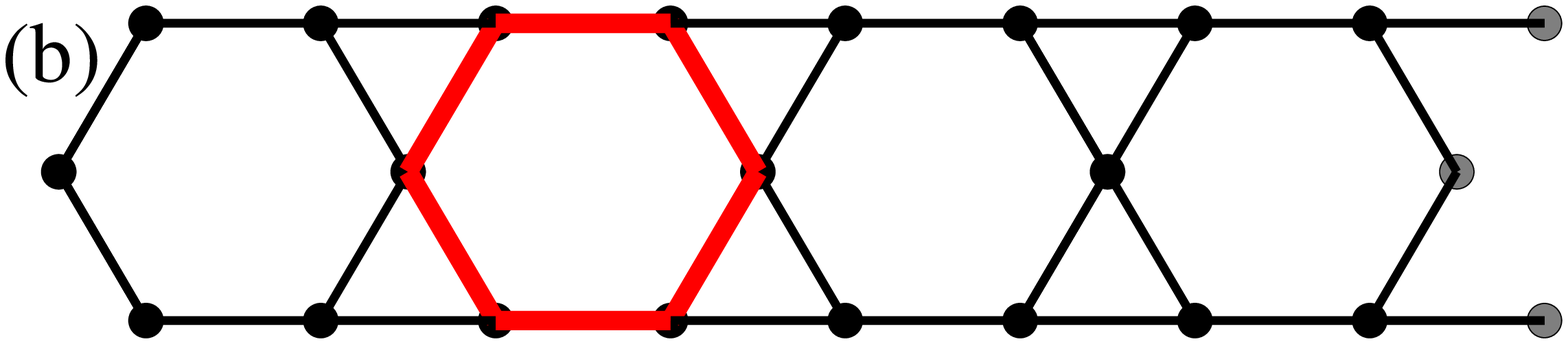}
\caption{\small
Kagome chain I (a) and kagome chain II (b).
Bold lines denote the minimal trapping cells for localized electrons.}
\label{lattices}
\end{center}
\end{figure}

In this paper we discuss ground-state magnetic properties of Mielke-type flat-band Hubbard ferromagnets
focusing for concreteness on two one-di\-men\-sional kagome lattices (Fig. \ref{lattices}).
The kagome chain I (panel a)
is a line graph of the two-leg ladder,
whereas the kagome chain II (panel b)
is a line graph of the decorated two-leg ladder \cite{mielke1,drhmm}.
The kagome chain I (II) has an elementary cell with 3 (5) sites.
We introduce the number of elementary cells ${\cal{N}}$
and ${\cal{N}}=N/3$ (${\cal{N}}=N/5$) for the kagome chain I (II).
For the even-${\cal{N}}$ kagome chain I and kagome chain II the parent graphs are bipartite ones,
whereas for the odd-${\cal{N}}$ kagome chain I the parent graph is nonbipartite
(periodic boundary conditions are implied).
Furthermore,
we consider the usual Hubbard Hamiltonian
\begin{eqnarray}
H=t\sum_{\sigma=\uparrow,\downarrow}\sum_{\langle i,j\rangle}(c_{i\sigma}^{\dagger}c_{j\sigma}+c_{j\sigma}^{\dagger}c_{i\sigma})
+U\sum_{i}n_{i\uparrow}n_{i\downarrow}.
\label{hamiltonian}
\end{eqnarray}
Here the sums run either over the nearest-neighbor pairs $\langle i,j\rangle$ or over $N$ lattice sites $i$.
$t>0$ is the hopping parameter
(then the flat one-electron band is the lowest-energy one)
and
$U>0$ is the on-site Coulomb repulsion for electrons with different spins.

We recall here Mielke's results \cite{mielke1} for ground-state ferromagnetism of the Hubbard model on line graphs
specifying them for the lattices in question.
The repulsive Hubbard model (\ref{hamiltonian}) has ferromagnetic ground states with saturated magnetization
if the number of electrons satisfies $n\le n_{\max}$
[$n_{\max}={\cal{N}}+1$ (periodic even-${\cal{N}}$ kagome chain I and periodic kagome chain II)
or
$n_{\max}={\cal{N}}$ (periodic odd-${\cal{N}}$ kagome chain I)]
and
the ferromagnetic ground state is unique if $n=n_{\max}$.
In what follows we illustrate the general statements of A.~Mielke and further specify them
1) constructing explicitly the ground states for $n\le n_{\max}$ electrons
and
2) calculating the average of the square of the total spin in these ground states $\langle {\bm{S}}^2 \rangle_n$.

We begin with introducing localized one-electron states.
A localized one-electron state can be located on a smallest diamond-shaped (hexagon-shaped) cell of the kagome chain I (II).
It is proportional to $\sum_m(-1)^mc_{m\sigma}^{\dagger}\vert 0\rangle$,
where the sum runs over 4 (6) sites of the diamond (hexagon) trapping cell of the kagome chain I (II).
The localized electron has the energy $-2t$
and it cannot escape from the trap for the given lattice topology
(isosceles triangles with one side belonging to the trap)
because of destructive quantum interference.
For the even-${\cal{N}}$ kagome chain I and the kagome chain II
there are two more traps for localized states,
i.e., the upper and lower legs.
Indeed, for periodic boundary conditions imposed
each leg is a regular convex polygon with an even number of sides/vertices
and $\sum_n(-1)^nc_{n\sigma}^{\dagger}\vert 0\rangle$, where the sum runs over the sites belonging to the leg,
is an eigenstate with the energy $-2t$.
Moreover,
isosceles triangles with one side belonging to the leg prevent escape of an electron from the leg.
The constructed ${\cal{N}}+2$ one-electron localized states with the energy $-2t$ obey one linear relation
and,
as a result,
we have only ${\cal{N}}+1$ linearly independent localized states in the subspace with $n=1$ electron.
Thus,
we have constructed $n_{\max}$
($n_{\max}={\cal{N}}$ for the periodic odd-${\cal{N}}$ kagome chain I
or
$n_{\max}={\cal{N}}+1$ for the periodic even-${\cal{N}}$ kagome chain I and kagome chain II)
one-electron localized states with the energy $-2t$.

We pass to subspaces with $1<n\le n_{\max}$ electrons.
Since the localized one-electron states are located within a restricted part of the lattice only,
we can construct many-electron eigenstates placing localized electrons
sufficiently far from each other.
Moreover,
the Hubbard repulsion is a positive semidefinite operator which can only increase the energy.
As a result,
the isolated localized electron states which are the ground states for $U=0$
remain the ground states for $U>0$.
Furthermore,
electrons obviously can be localized within neighboring trapping cells if they all have the same spin polarization,
since in such a case the on-site Hubbard repulsion is irrelevant.
Starting from such a spin-polarized state and using the global SU(2) invariance of the Hubbard model (\ref{hamiltonian})
more ground states can be trivially constructed.
Denote by $g_{{\cal{N}}}(n)$ the number of the ground states in the subspace with $n\le n_{\max}$ electrons.
The ground states for the odd-${\cal{N}}$ kagome chains I
are constructed from the one-electron states trapped by diamonds only.
Repeating the arguments elaborated for the sawtooth Hubbard chain
(see Ref.\ \onlinecite{dhr07}a)
we arrive at the following result:
$g_{{\cal{N}}}(n)=D_{{\cal{N}}}(n)$,
$D_{{\cal{N}}}(n)={\cal{Z}}(n,{\cal{N}})+({\cal{N}}-1)\delta_{n,{\cal{N}}}$,
where ${\cal{Z}}(n,{\cal{N}})$ is the canonical partition function
of $n$ classical hard dimers
on an auxiliary (periodic) chain of $2{\cal{N}}$ sites.
For the even-${\cal{N}}$ kagome chains I and kagome chains II
the states which are constructed also from the one-electron leg states
are among the ground states for $n=1,\ldots,{\cal{N}}+1$
and, as a result,
$g_{{\cal{N}}}(n)=D_{{\cal{N}}}(n)(1-\delta_{n,{\cal{N}}+1})+L_{{\cal{N}}}(n)(1-\delta_{n,0})$
with $L_{{\cal{N}}}(n)=(n+1) {{{\cal{N}}}\choose{n-1}} +\delta_{n,2}$.
Further details can be found in Ref.\ \onlinecite{drhmm}.

To reveal the ferromagnetic ground states
we consider the equal-weight average over all $g_{{\cal{N}}}(n)$ ground states for a given number of electrons $n$ of the total spin,
$\langle {\bm{S}}^2\rangle_n=3\langle {S^z}^2\rangle_n$.
Since each of the $D_{{\cal{N}}}(n)$ states can be encoded
by a certain configuration of $n$ hard dimers on the chain of $2{\cal{N}}$ sites \cite{dhr07,drhmm},
$\langle{S^z}^2\rangle_n$ is related to density-density correlation functions of the hard-dimer system.
For the even-${\cal{N}}$ kagome chain I or kagome chain II
we have to take into account in addition the ground states involving the leg states
all of which are ferromagnetic (except one state for $n=2$ electrons).
In Table \ref{table1} we report $\langle {\bm{S}}^2\rangle_n$ for the two kagome chains with 7 cells.
\begin{table}[t!]
\caption{\label{table1}
$\langle {\bm{S}}^2\rangle_n$, $n=1,\ldots,n_{\max}$ for the kagome chains I and II with ${\cal{N}}=7$ cells.
Localized-state predictions coincide with exact diagonalization data for the Hubbard model.}
\begin{center}
\begin{tabular}{ccccccccc}
\hline
   &$n=1$    &$n=2$    &$n=3$    &$n=4$    &$n=5$    &$n=6$    & $n=7$   &$n=8$\\
\hline
\hline
I  &$\frac{3}{4}$    &$\frac{18}{11}$  &$\frac{11}{4}$   &$\frac{30}{7}$   &$\frac{27}{4}$   &$12$     &$\frac{63}{4}$   &     \\
II &$\frac{3}{4}$    &$\frac{83}{49}$  &$\frac{85}{28}$  &$\frac{330}{67}$ &$\frac{903}{116}$&$12$     &$\frac{63}{4}$   &$20$ \\
\hline
\end{tabular}
\end{center}
\end{table}
Using a MAPLE code
we have computed $\langle {\bm{S}}^2\rangle_n$, $n=1,\ldots,n_{\max}$
for essentially longer chains with several hundred cells \cite{drhmm}.
Examining $\langle {\bm{S}}^2\rangle_n/{\cal{N}}^2$ vs $n/{\cal{N}}$ as ${\cal{N}}$ increases
we have found that there is no finite region of electron density $n/{\cal{N}}<1$
for which $\langle {\bm{S}}^2\rangle_n/{\cal{N}}^2$ remains finite for both kagome chains as ${\cal{N}}\to\infty$
(in this limit $\langle {\bm{S}}^2\rangle_n\to 3n/4$) \cite{drhmm}.

Finally, one can add to the Hamiltonian (\ref{hamiltonian}) the Zeeman term
$-h\sum_{i}(c^{\dagger}_{i\uparrow}c_{i\uparrow}-c^{\dagger}_{i\downarrow}c_{i\downarrow})/2$,
which removes the degeneracy of SU(2) multiplets. We have also calculated
the localized-electron contribution to the partition function in the
presence of the Zeeman term and determined the magnetization and
susceptibility, confirming in an alternative way the results of the
preceding paragraph.

In summary,
we have illustrated in detail Mielke's results about ground-state ferromagnetism of the Hubbard model on line graphs
for the two kagome stripes.
An ambitious task is to extend such an analysis for the (two-dimensional) kagome lattice.

{\bf Acknowledgments:}
O.~D.\ thanks the MPIPKS-Dres\-den for kind hospitality in 2009
and during the workshop on Perspectives in Highly Frustrated Magnetism (April 19-23, 2010).
A.~H.\ acknowledges financial support by the DFG through a Heisenberg fellowship (project HO 2325/4-1).

\end{document}